\documentclass[iop]{emulateapj}
\usepackage{apjfonts}

\begin{document}

\epsscale{1.1}

\title{The Primeval Populations of the Ultra-Faint Dwarf 
Galaxies\altaffilmark{1}}
\shorttitle{The Primeval Populations of the UFD Galaxies}

\author{
Thomas M. Brown\altaffilmark{2}, 
Jason Tumlinson\altaffilmark{2}, 
Marla Geha\altaffilmark{3}, 
Evan N. Kirby\altaffilmark{4,5}, 
Don A. VandenBerg\altaffilmark{6},\\ 
Ricardo R. Mu\~noz\altaffilmark{7}, 
Jason S. Kalirai\altaffilmark{2}, 
Joshua D. Simon\altaffilmark{8}, 
Roberto J. Avila\altaffilmark{2},\\ 
Puragra Guhathakurta\altaffilmark{9}, 
Alvio Renzini\altaffilmark{10}, and
Henry C. Ferguson\altaffilmark{2} 
}

\altaffiltext{1}{Based on observations made with the NASA/ESA {\it Hubble
Space Telescope}, obtained at STScI, which
is operated by AURA, Inc., under NASA contract NAS 5-26555.}

\altaffiltext{2}{Space Telescope Science Institute, 3700 San Martin Drive,
Baltimore, MD 21218, USA;  
tbrown@stsci.edu, tumlinson@stsci.edu, jkalirai@stsci.edu, avila@stsci.edu,
ferguson@stsci.edu}

\altaffiltext{3}{Astronomy Department, Yale University, New Haven, CT
06520, USA; marla.geha@yale.edu}

\altaffiltext{4}{California Institute of Technology, 1200 East California 
Boulevard, MC 249-17, Pasadena, CA 91125, USA; enk@astro.caltech.edu}

\altaffiltext{5}{Hubble Fellow}

\altaffiltext{6}{Department of Physics and Astronomy, 
University of Victoria, P.O. Box 3055, Victoria, BC, V8W 3P6, Canada; 
vandenbe@uvic.ca}

\altaffiltext{7}{Departamento de Astronom\'ia, Universidad de Chile, 
Casilla 36-D, Santiago, Chile; rmunoz@das.uchile.cl}

\altaffiltext{8}{Observatories of the Carnegie Institution of Washington, 
813 Santa Barbara Street, Pasadena, CA 91101, 
USA; jsimon@obs.carnegiescience.edu}

\altaffiltext{9}{UCO/Lick Observatory and Department of Astronomy and 
Astrophysics, University of California, Santa Cruz, CA 95064, USA; 
raja@ucolick.org}

\altaffiltext{10}{Osservatorio Astronomico, Vicolo Dell'Osservatorio 5, 
I-35122 Padova, Italy; alvio.renzini@oapd.inaf.it}

\submitted{Accepted for publication in The Astrophysical Journal Letters}

\begin{abstract}

We present new constraints on the star formation histories of the
ultra-faint dwarf (UFD) galaxies, using deep photometry obtained with
the {\it Hubble Space Telescope (HST)}.  A galaxy class recently
discovered in the Sloan Digital Sky Survey, the UFDs appear to
be an extension of the classical dwarf spheroidals to low
luminosities, offering a new front in efforts to understand the
missing satellite problem.  They are the least luminous, most
dark-matter dominated, and least chemically-evolved galaxies known.
Our {\it HST} survey of six UFDs seeks to determine if these galaxies
are true fossils from the early universe.  We present here the
preliminary analysis of three UFD galaxies: Hercules, Leo~IV, and Ursa
Major~I.  Classical dwarf spheroidals of the Local Group exhibit
extended star formation histories, but these three Milky Way
satellites are at least as old as the ancient globular cluster M92,
with no evidence for intermediate-age populations.  Their ages also
appear to be synchronized to within $\sim$1~Gyr of each other, as
might be expected if their star formation was truncated by a global
event, such as reionization.

\end{abstract}


\section{Introduction}

Although the Lambda Cold Dark Matter paradigm is consistent with many
observable phenomena, 
discrepancies arise at small scales.  One of the most prominent issues
is that it predicts many more dark-matter halos than are actually seen
as dwarf galaxies (e.g., Moore et al.\ 1999).  A possible solution has
arisen with the recent discovery of additional satellites around the
Milky Way (e.g., Willman et al.\ 2005; Zucker et al.\ 2006; Belokurov
et al.\ 2007) and Andromeda (e.g., Zucker et al.\ 2007) in the
Sloan Digital Sky Survey (York et al.\ 2000) and other
wide-field surveys (e.g., McConnachie et al.\ 2009).  

The newly-discovered ultra-faint dwarf (UFD) galaxies
appear to be an extension of the classical dwarf spheroidals (dSphs)
to lower luminosities ($M_V \gtrsim -8$~mag).
UFD luminosities are comparable to those of globular clusters, but one
distinction in the former is the presence of dark
matter.  Even massive globular clusters have mass-to-light ratios
($M/L_V$) of $\sim$2 (e.g., Baumgardt et al.\ 2009; van de Ven
et al.\ 2006), precluding significant dark matter.  In contrast, all
known dwarf galaxies have higher $M/L_V$ (Kalirai et al.\ 2010
and references therein). UFD kinematics are clearly dark
matter-dominated, with $M/L_V > 100$ (e.g., Kleyna et al.\ 2005; Simon
\& Geha 2007; Mu\~noz et al.\ 2006), even where velocity dispersions have been
revised downward (e.g., Koposov et al.\ 2011).
The inferred dark-matter
densities of dwarf galaxies suggest a high-redshift collapse for both
classical dSphs and UFDs ($z \sim 12$; Strigari et al.\ 2008), but the
dSphs apparently continued to evolve (Orban et al.\ 2008; Weisz et
al.\ 2011).  In contrast, the UFDs are the least chemically-evolved
galaxies known, with abundance patterns that imply their star
formation was brief (Frebel et al.\ 2010) 
and individual stellar metallicities as low as
[Fe/H]~=~$-3.7$ (Norris et al.\ 2010). 
The strict conformance to a metallicity-luminosity relation for all
Milky Way satellites limits the amount
of tidal stripping to a factor of $\sim$3 in stellar mass (Kirby et
al. 2011).  Therefore, UFDs are not tidally stripped versions of
classical dSphs (see also Penarrubia et al. 2008; Norris et al.\ 2010).

\begin{table*}[t]
\begin{center}
\caption{Observations}
\begin{tabular}{ccccccccccccc}
\tableline
             &         &         &          &        &      &                             &                       &     & \multicolumn{2}{c}{Exposure per tile} & \multicolumn{2}{c}{50\% complete}\\
             & R.A.    & Dec.    & $(m-M)_V$&$E(B-V)$&$M_V$ &                             &[Fe/H]                 &     & F606W      & F814W                    & F606W & F814W \\
Name         & (J2000) & (J2000) & (mag)    &(mag)   &(mag) &$<$[Fe/H]$>$\tablenotemark{a}&r.m.s.\tablenotemark{b}&tiles&  (s)       & (s)                      & (mag) & (mag) \\
\tableline
Hercules     & 16:31:05&+12:47:07& 20.92$\pm$0.06    &0.08$\pm$0.02    &$-6.2$\tablenotemark{c}&$-2.41$                      &0.6                    & 2   & 12,880     & 12,745  & 29.1  & 29.1\\
Leo IV       & 11:32:57&-00:31:00& 21.15$\pm$0.08    &0.05$\pm$0.02    &$-5.8$\tablenotemark{d}&$-2.54$                      &0.9                    & 1   & 20,530     & 20,530  & 29.1  & 29.2\\
Ursa Major I & 10:35:04&+51:56:51& 20.11$\pm$0.04    &0.04$\pm$0.02    &$-5.5$\tablenotemark{e}&$-2.18$                      &0.7                    & 9   &  4,215     &  3,725  & 28.4  & 28.4\\
\tableline
\end{tabular}
\end{center}
\tablenotetext{a}{Kirby et al.\ (2011), 
based on Simon \& Geha et al.\ (2007) spectroscopy.}
\tablenotetext{b}{For stars with $<$0.3~dex uncertainties.}
\tablenotetext{c}{Sand et al.\ (2009)}
\tablenotetext{d}{de Jong et al.\ (2010)}
\tablenotetext{e}{Martin et al.\ (2008)}
\end{table*}

As one way of solving the missing satellite problem, 
galaxy formation simulations assume that UFDs formed the bulk
of their stars prior to the epoch of reionization (e.g., Tumlinson
2010; Mu\~noz et al.\ 2009; Bovill \& Ricotti 2009; Koposov
et al.\ 2009).  Mechanisms that could drive an early
termination of star formation include reionization, gas depletion, and
supernova feedback.  Using the {\it Hubble Space Telescope (HST)}, we are
undertaking a deep imaging survey of UFDs that reaches the old
main sequence (MS) in each galaxy, yielding high-precision
color-magnitude diagrams (CMDs) that provide sensitive probes of their
star formation histories.  The program includes Hercules, Leo~IV,
Ursa Major~I, Bootes~I, Coma Berenices, and Canes Venatici~II.  
Here, we give preliminary results for the first three galaxies.

\section{Observations and Data Reduction}

We obtained deep optical images of each galaxy (Table~1) using the
F606W and F814W filters on the Advanced Camera for Surveys (ACS).
These filters efficiently enable a high signal-to-noise ratio (SNR) on
the stellar MS and facilitate comparison with other {\it HST} programs
exploring the Local Group.  The image area and depth were chosen to provide
a few hundred stars on the upper MS, with a SNR of $\sim$100 near the
turnoff.  Our image processing includes the latest pixel-based
correction (Anderson \& Bedin 2010; Anderson 2012, in prep.) 
for charge-transfer inefficiency (CTI).  All images were
dithered to mitigate detector artifacts and enable resampling of the
point spread function (PSF).  We co-added the images for each filter
in a given tile using the IRAF {\sc drizzle} package (Fruchter \& Hook
2002), including masks for cosmic rays and hot pixels,
resulting in geometrically-correct images with a plate
scale of 0.03$^{\prime\prime}$ pixel$^{-1}$ and an area of
approximately $210^{\prime\prime} \times 220^{\prime\prime}$.

We performed both aperture and PSF-fitting photometry using the
DAOPHOT-II package (Stetson 1987), assuming a spatially-variable PSF
constructed from isolated stars.  The final catalog combined aperture
photometry for stars with photometric errors $<$0.02~mag and
PSF-fitting photometry for the rest, all normalized to
an infinite aperture.  Due to the scarcity of bright stars, the uncertainty
in the normalization is 0.02~mag.  Our
photometry is in the STMAG system: $m= -2.5 \times $~log$_{10}
f_\lambda -21.1$.  For galaxies with multiple tiles, the ridge line in
each tile was compared to that in other tiles; we then made small
($\sim$0.01~mag) color adjustments to force agreement between tiles,
ensuring the full catalog for a galaxy did not have larger photometric
scatter than that in any tile.  The catalogs were cleaned
of background galaxies and stars with poor photometry,
rejecting outliers in $\chi^2$ (from PSF fitting), PSF sharpness,
and photometric error.  We also rejected stars with bright neighbors
or falling within extended background galaxies.  We performed
artificial star tests to evaluate photometric scatter and
completeness, including CTI effects, with the same photometric
routines used to create the photometric catalogs.

\begin{figure*}[t]
\plotone{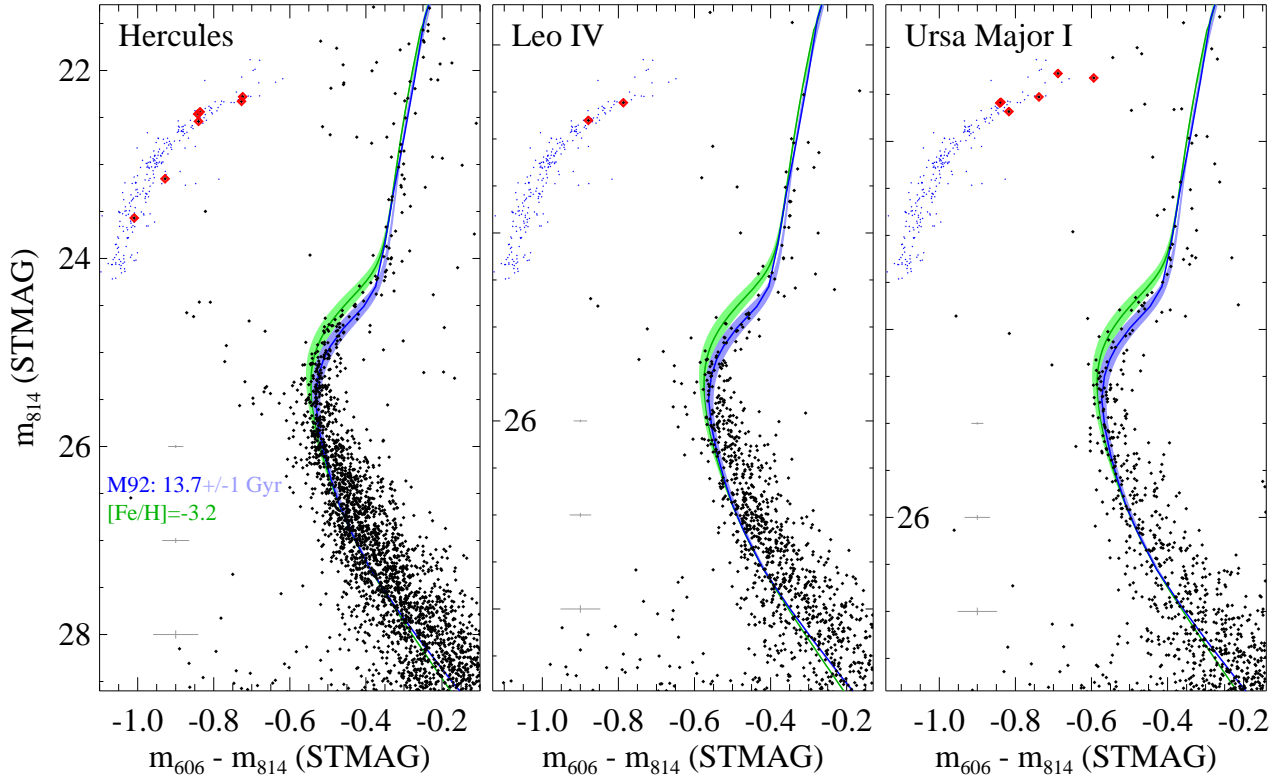}
\caption{The CMD of each UFD (black points), with axes shifted to ease
  comparisons.  For reference, we show
  the empirical ridge line for the MS, SGB, and RGB in M92 (dark blue
  curve), along with the HB locus in M92 (blue points).  The M92
  fiducial has been placed at the distance and reddening for
  each galaxy, matching the luminosity of HB stars 
  (highlighted in red) and the color of the lower RGB stars in each
  UFD (accounting for the metallicity distribution).
  The light blue band is bounded
  by isochrones within 1 Gyr of M92's age (13.7 Gyr), at the M92
  metallicity ([Fe/H]=$-2.3$), showing excellent agreement with the
  M92 ridge line.  Because the UFD metallicities extend much
  lower than those in globular clusters, we also show an isochrone at
  the age of M92 but at the low extreme of the metallicity
  distribution ([Fe/H]=$-3.2$; dark green curve), bounded by
  isochrones offset by 1~Gyr (light green band).  A
  blue straggler sequence is present in each galaxy, 
  but there are no stars significantly younger than M92.}
\end{figure*}

The CMD of each UFD galaxy is shown in Figure~1.  Each UFD is clearly
dominated by an ancient population.  Field contamination is
significant, but is particularly high in the vicinity of the red giant
branch (RGB).  The expected number of RGB stars can be estimated from
the number of subgiant branch (SGB) stars, given the relative
lifetimes implied by isochrones. The implied contamination varies from
roughly 30\% at 1~mag above the SGB to twice that at 2~mag above the
SGB, where this contamination increases the apparent RGB breadth.  In
Hercules and Leo~IV, the star counts are a bit lower than expectations
from ground surveys (Mu\~noz et al.\ in prep.), while the counts in
Ursa Major~I are significantly lower, but the sample is large enough
to judge its population relative to those of the other two galaxies.
\\ \\

\section{Analysis}

\subsection{Comparison to M92}

As one of the most ancient, metal-poor, and well-studied globular
clusters, M92 is an important empirical fiducial for the UFD galaxies.
It was previously observed by Brown et al.\ (2005) with ACS, using the
same filters employed here, as one of six empirical templates for deep
CMD analyses of Andromeda (e.g., Brown et al.\ 2006).  Since then,
estimates of the M92 distance modulus have generally drifted
$\sim$0.1~mag higher, and estimates for the metallicity have generally
decreased by $\sim$0.2 dex.  We assume $(m-M)_{\rm o} = 14.62$~mag, which is
the mean of Paust et al.\ (2007; 14.60$\pm$0.09~mag), Del Principe et
al.\ (2005; 14.62$\pm$0.1~mag), and Sollima et al.\ (2006;
14.65$\pm$0.1~mag).  We assume $E(B-V)$~=~0.023~mag (Schlegel et
al.\ 1998), and [Fe/H]~=~$-2.3$ (Harris et al.\ 1996).  Because the
UFD metallicity distribution extends much lower than that of any
globular cluster, the age determination must rely upon theoretical
isochrones (see \S3.2), but a comparison to M92 is still instructive.

In our analysis, the largest uncertainties are the distances of the
UFDs relative to the globular cluster calibrators of our isochrone
library, with emphasis on M92.  We determined apparent distance
moduli by minimizing the luminosity differences between the UFD
horizontal branch (HB) stars (Figure 1; red points) and the horizontal
region of the M92 HB locus (blue points).  We determined the
foreground extinction by minimizing the color differences between UFD
stars on the lower RGB and the RGB color distribution implied by the
spectroscopic metallicties associated with each galaxy (Simon \& Geha
2007; Kirby et al.\ 2008), anchored to M92 (which is near the mean
metallicity of each UFD).  The uncertainties in our determinations of
distance and reddening are from the standard deviation in Monte Carlo
realizations of these fits, excluding any uncertainties
in the distance and reddening of M92.  We show our derived apparent
distance moduli and reddening values in Table~1, but note comparisons
to the recent literature here.  For Hercules, Sand et al.\ (2009)
found $(m-M)_{\rm o} = $~20.625$\pm$0.1~mag and assumed $E(B-V) =
0.064$~mag (Schlegel et al.\ 1998), implying $(m-M)_V =$~20.8~mag.
For Leo~IV, Moretti et al.\ (2009) found $(m-M)_{\rm o}
=$~20.94$\pm$0.07~mag and $E(B-V) =$~0.04$\pm$0.01~mag using RR Lyrae
stars (associated with an old population), implying $(m-M)_V
=$~21.06~mag.  For Ursa Major~I, Okamoto et al.\ (2008) found
$(m-M)_{\rm o}=$~19.93$\pm$0.1~mag and assumed $E(B-V) =$0.019~mag (Schlegel
et al.\ 1998), implying $(m-M)_V =$~19.99~mag.  Note that if we
were to assume the shorter distances and/or lower reddening in the
literature, the UFD ages relative to M92 would {\it increase}.  The
observed UFD CMDs would not change, but the M92 ridge line would shift
brighter and bluer, making the MS turnoff in each galaxy appear
fainter and redder (and thus older) relative to that in M92, which
would significantly violate the age of the universe ($13.75 \pm 0.11$;
Jarosik et al.\ 2011) unless other factors (e.g., CNO abundances) were
adjusted accordingly.

With these caveats, we show the comparison of each UFD to M92 in
Figure~1.  HB morphology is most sensitive to metallicity (but also
age and the abundances of He and CNO), with the HB color distribution
shifting toward the blue at lower [Fe/H].  The colors of the HB stars
in each UFD are consistent with their metallicities, in the sense that
$<$[Fe/H]$>$ is slightly higher in Ursa Major~I than in Hercules or
Leo~IV (Table~1), and its HB stars are somewhat redder, but this may
be due to small number statistics.  The RGB appears
slightly wider than one would expect from the photometric errors and
metallicity distribution, but this is uncertain, given the field
contamination.  The MS turnoff and SGB in each UFD
extends slightly brighter and bluer than those features on the M92
ridge line.  If we did not know the metallicity distributions, we
would conclude that the brighter and bluer stars have ages a few Gyr
younger or metallicities up to $\sim$1~dex lower than M92.  However,
we know the UFD metallicities extend below [Fe/H]~=~$-3$, so these
stars are consistent in age with M92, as we show below.

\begin{figure*}[t]
\plotone{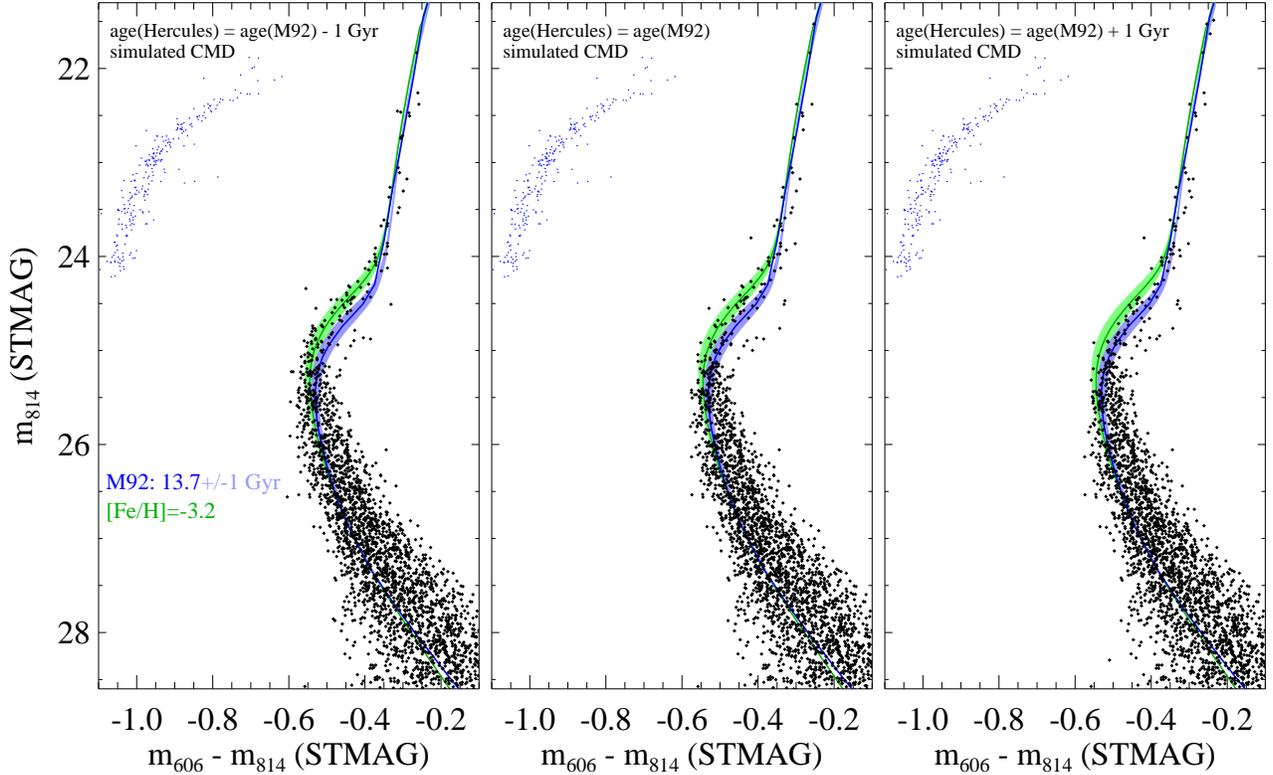}
\caption{Simulated CMDs for Hercules (MS--RGB) under three assumptions: 
a population 1~Gyr younger than M92 (left panel), a population at the
age of M92 (middle panel), and a population 1~Gyr older than M92 (right 
panel).  To ease comparison with the observed Hercules CMD, these simulations
are plotted on the same scale as Figure 1 and include the same empirical and 
theoretical fiducials (colored curves).  The middle panel best matches
the observed Hercules CMD, given the luminosity distribution of the SGB
stars and the color of the MS turnoff (bluest point on the MS).}
\end{figure*}

\subsection{Comparison to Theoretical Isochrones}

Because the UFDs are among the most metal-poor populations in the local
universe, we cannot rely solely upon empirical stellar population
templates (e.g., globular clusters) to analyze them.
We must extend these empirical templates to lower metallicities
using theoretical models.  For this purpose, we employ the
Victoria-Regina isochrones (VandenBerg et al.\ 2012).

Compared to the previous version of the code (VandenBerg et
al.\ 2006), the current version includes, among other things, the
effects of He diffusion, the latest improvements to the H-burning
nuclear reaction rates, and an update to the Asplund et al.\ (2009)
solar metals mixture.  The net effect of these changes is to reduce
the age at a given turnoff luminosity by $\sim$0.5~Gyr.  Our purpose
here is to use these isochrones to derives ages {\it relative} to
those of the globular clusters, but these effects should be kept in mind
when considering absolute ages.

Brown et al.\ (2005) produced a set of empirical population templates
in the ACS F606W and F814W filters, spanning a wide metallicity range,
based upon Galactic globular and old open clusters.  They then
calibrated a transformation of the Victoria-Regina isochrones into the
same filters, producing agreement at the 1\% level.  Here, we
have adjusted this transformation by 0.03~mag in color.
This adjustment accounts for the new version of the Victoria-Regina
isochrones and the current state of the literature regarding cluster
parameters.  These isochrones match the ACS CMDs of 47~Tuc at an age
of 12.3 Gyr, NGC~6752 at an age of 13.2~Gyr, and M92 at an age of
13.7~Gyr.  As discussed in \S3.1, M92 is the most
relevant calibrator, but the other clusters demonstrate that the
isochrones are valid over a wide range of [Fe/H]
($-0.7$ to $-2.3$).  We stress that these absolute ages depend critically 
upon the assumed distance moduli and oxygen abundances.

In Figure~1, we show the expected location of stars in the ACS CMDs
for ages within a Gyr of the M92 age (i.e., 12.7--14.7 Gyr) at two
different metallicities: that of M92 (i.e., [Fe/H]$=-2.3$; blue curve)
and the low extreme of the metallicity distribution in these UFDs
([Fe/H]$=-3.2$; green curve).  It is clear from the comparison that
the luminosity of the MS turnoff and SGB in each galaxy is consistent
with ancient metal-poor stars.  The few stars
immediately brighter and bluer than the old MS turnoff are blue
stragglers, which are ubiquitous in old populations.  If these were
young stars, each UFD would have a star-formation history consisting of
two delta functions in age, with a few percent of the stars 
at $\sim$4~Gyr and the rest ancient; continuous
low-level star formation would produce intermediate-age MS and SGB stars 
that are not observed.

Although the MS turnoff becomes bluer and brighter at both younger
ages and lower metallicities, this age-metallicity degeneracy can be
broken by simultaneously fitting the MS, SGB, and RGB in a CMD.
However, the scarcity of RGB stars and multitude of field stars
precludes this type of fit here.  Instead, we rely upon the observed
metallicity distribution from spectral synthesis of neutral Fe lines
(Kirby et al.\ 2008; based upon Keck spectroscopy from Simon \& Geha
2007).  For the current analysis, we will use the relatively
well-populated CMD of Hercules as an example, employing the 15 RGB
stars with low spectroscopic errors ($<$0.3~dex) observed to date.  We
constructed a library of CMD probability clouds representing simple
stellar populations spanning $-1.4 >$[Fe/H]$>-3.2$ and ages of
10--16~Gyr.  The probability clouds were populated using the
Victoria-Regina isochrones, a Salpeter IMF, and the photometric
scattering and completeness from the artificial star tests.  Synthetic
CMDs constructed from this library were fit to the observed CMD by
allowing the ages of the components to float while holding the
metallicity distribution fixed, using a downhill simplex to minimize
the Maximum Likelihood statistic of Dolphin (2002).  The fits were
repeated with a fine grid of starting conditions to avoid a spurious result
driven by local minima in the fit.
Uncertainties in the derived age were determined by fitting repeated
Monte Carlo realizations of the observed CMD.  The CMD was fit in the
vicinity of the MS turnoff ($24.0 < m_{F814W} < 25.7$~mag), given its
sensitivity to age, while excluding the lower MS stars that provide no
age leverage.  Note that the mass range spanned by our fit is only
$\sim$0.03 $M_\odot$, and thus insensitive to the assumed IMF.  We find
a mean age of 13.6~Gyr for Hercules.  Formally, the statistical
uncertainty is 0.2 Gyr, consistent with the age of M92
(13.7~Gyr).  The systematic uncertainties from distance and [O/Fe]
are each potentially as large as 0.6~Gyr.

To demonstrate the sensitivity of these CMDs to age, we show in
Figure~2 three simulated CMDs for Hercules.
Each CMD assumes the observed spectroscopic
metallicity distribution, under three age assumptions: 1
Gyr younger than M92 (left panel), the same age as M92 (middle panel),
and 1 Gyr older than M92 (right panel).  It is clear that the best
match to the observed Hercules CMD (Figure~1) is the middle panel,
where the populations have the same age as M92.  Although the
population statistics are best in Hercules, the similarity of the
Leo~IV and Ursa Major~I CMDs implies that they also host populations
of approximately the same age.  None of these galaxies appears to host
a significant population of stars younger than M92.

In the Hercules spectroscopy, approximately 20\% of the RGB stars fall
at [Fe/H]~=~$-1.7 \pm 0.3$.  In the Hercules CMD (Figure~1), the
stars immediately below the M92 SGB, especially near the MS, are
consistent with such stars at ages within a Gyr of M92 (cf.\ Figure
2), although the exact number is uncertain, given the field
contamination.  However, if any of these stars at [Fe/H]$> -2$ were as
much as 2 Gyr younger than M92, they would be difficult to discern in
the Hercules CMD, because the SGB for such stars would fall within the
dominant SGB observed for Hercules (i.e., within the region bounded by
the blue and green fiducials of Figure 1).  Thus, while it is clear
that most stars in Hercules are as old as those in M92, we cannot at
this time rule out a minority sub-population of stars (up to
$\sim$10\% of the total population) that are 1--2~Gyr younger at the
high end of the Hercules metallicity distribution.

\section{Discussion}

The three UFD galaxies here are among the more distant of those known
in the Milky Way system, at 100--150~kpc, so ground-based CMDs of each
implied they were old ($>$10~Gyr), but could not put tight constraints
on their ages (e.g., Sand et al.\ 2009; Sand et al.\ 2010; Okamoto et
al.\ 2008, 2012; Ad\'en et al.\ 2010).  Our {\it HST} observations
reach well below the MS turnoff in each galaxy, revealing that all
three host truly ancient metal-poor populations.  The majority of the
stars in each galaxy must have ages within 1~Gyr of M92's age, with
younger ages strongly ruled out, although we cannot exclude a trace
population of stars 1--2~Gyr younger than M92 at the high end of the
metallicity distribution.

Two UFD galaxies (Bootes~I and Coma Berenices) are significantly
closer (44--66 kpc), such that ground-based CMDs can place constraints
on their ages approaching those possible with {\it HST}
(Mu\~noz et al.\ 2010; Okamoto et al.\ 2012).  These CMDs also imply
ages approximately as old as M92, although the use of distinct
bandpasses hampers accurate comparisons to our observations.
Our {\it HST} survey includes these galaxies, enabling
accurate age measurements in a significant sample of UFDs, with all
observed in the same photometric system as each other and the most
ancient globular clusters.

If we include Bootes~I and Coma Berenices, 
it seems likely that at least 5 UFD galaxies have ages consistent with
that of the oldest known globular cluster (M92), with no evidence for
significantly younger populations.  This is in striking
contrast to any other galaxy class in the local universe.
Our external vantage point for Andromeda enables accurate ages
throughout its halo, using the same instrument and techniques
described here, yet all such measurements to date have found an
extended star formation history, with significant numbers of stars
younger than 10~Gyr (Brown et al.\ 2006, 2008).  An {\it HST}
survey of 60 dwarf galaxies within 4~Mpc found that most formed the
bulk of their stars prior to $z \sim 1$, but none were consistent with a
purely ancient population (Weisz et al.\ 2011).  The UFDs may be the
only galaxies where star formation ended in the earliest epoch of the
universe.  If so, the apparent synchronicity to their star formation
histories suggests a truncation induced by a global event, such
as reionization (13.3~Gyr ago; Jarosik et al.\ 2011).
These UFDs were likely the victims of reionization, rather than the 
agents, given the small numbers of stars available
to produce ionizing photons.  

\acknowledgements

Support for GO-12549 was provided by NASA through a grant from
STScI, which is operated by AURA, Inc., under NASA contract NAS
5-26555.  ENK acknowledges support by NASA through Hubble Fellowship
grant 51256.01 from STScI.  AR acknowledges support from ASI via grant
I/009/10/0.  R.R.M. acknowledges support from the GEMINI-CONICYT Fund,
allocated to the project N$^{\circ}32080010$.

\end{document}